\documentclass{article}
\usepackage{graphicx}

\begin{document}

\title{
Lotka-Volterra population model of genetic evolution
}

\author{Miros{\l}aw R. Dudek\footnote{mdudek@proton.if.uz.zgora.pl},\\
Institute of Physics, Zielona G{\'o}ra University,\\
65-069 Zielona G{\'o}ra, Poland 
}

\maketitle
\date{}



\begin{abstract}

A deterministic model of an age-structured
population with genetics analogous to the  discrete time Penna model
\cite{Penna1,Penna2} of genetic evolution is constructed on the basis
of the Lotka-Volterra scheme. It is shown that if, as in the Penna model, 
genetic information is represented by the fraction of defective genes 
in the population, the population numbers for each specific individual's 
age are represented by exactly the same functions of age in both models. 
This gives us a new possibility to
consider multi-species evolution without using detailed microscopic
Penna model.

We discuss a particular case of the predator-prey system representing
an ecosystem consisting of a limited amount of energy resources
consumed by the age-structured species living in this ecosystem. Then,
the increase in number of the individuals in the population under
consideration depends on the available energy resources, the shape of
the distribution function of
  defective genes in the population and the fertility 
age. We show that these parameters determine the trend toward
equilibrium of the whole ecosystem. 
\end{abstract}

\section{Introduction}

\begin{figure}[t]
\begin{center}
\includegraphics[scale=0.5]{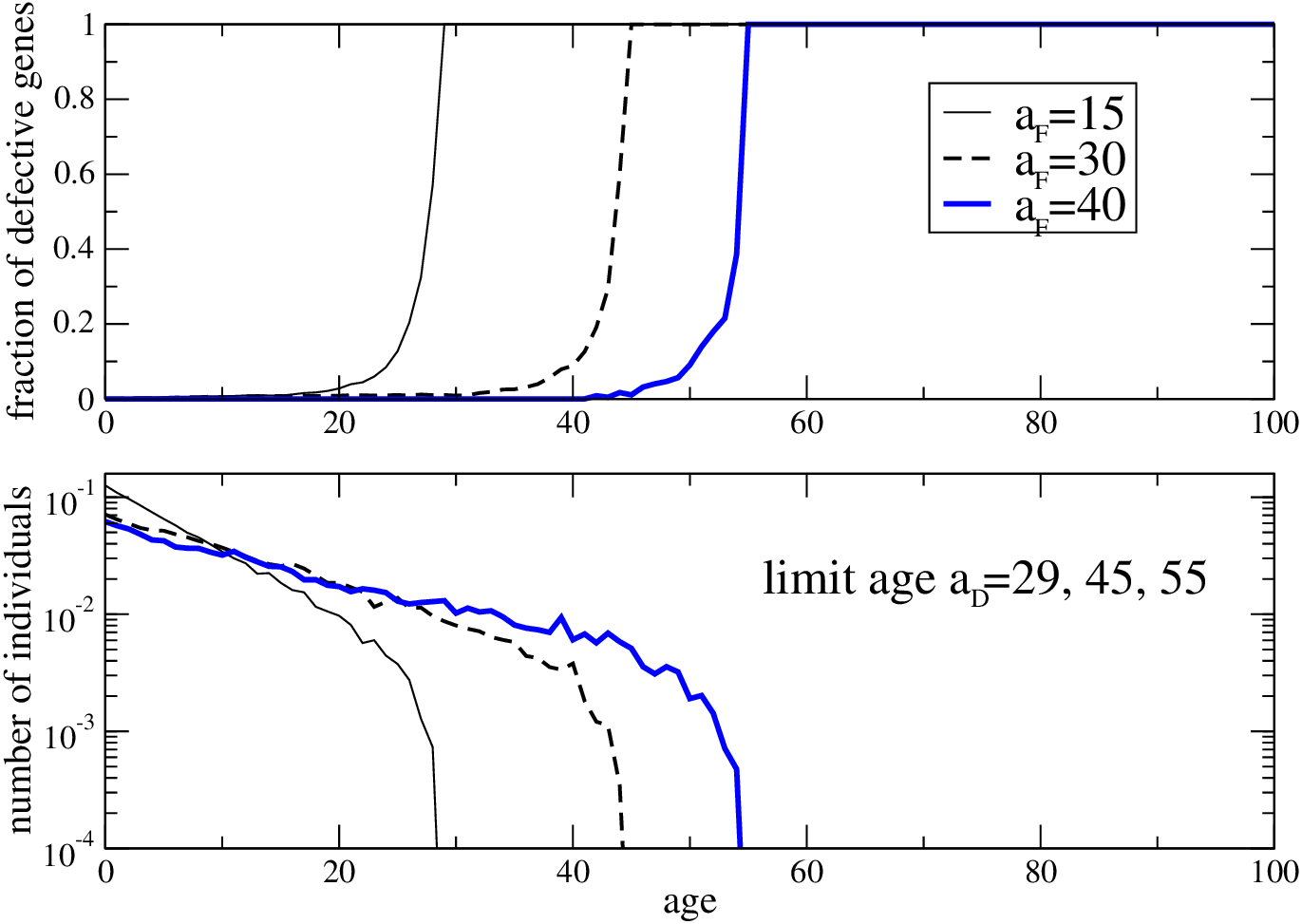}
\end{center}
\caption{Monte Carlo simulation results of the Penna model of genetic evolution after $5 \times 10^7$ MC steps. The upper panel shows the age-specific fraction of defective genes 
in three haploid populations with different values of the individual's fertility age $a_F=15, 30, 40$. The remaining parameters of the Penna model are the same, $T=1$, $M=1$, bit-string length $n=100$. The resulting limit age of individuals in these populations $a_D=29, 45, 55$ is the age where the fraction of defective genes is equal to 1.
The lower panel shows the age distribution in these populations. 
}
\label{fig1}
\end{figure}

\begin{figure}[t]
\begin{center}
\includegraphics[scale=0.5]{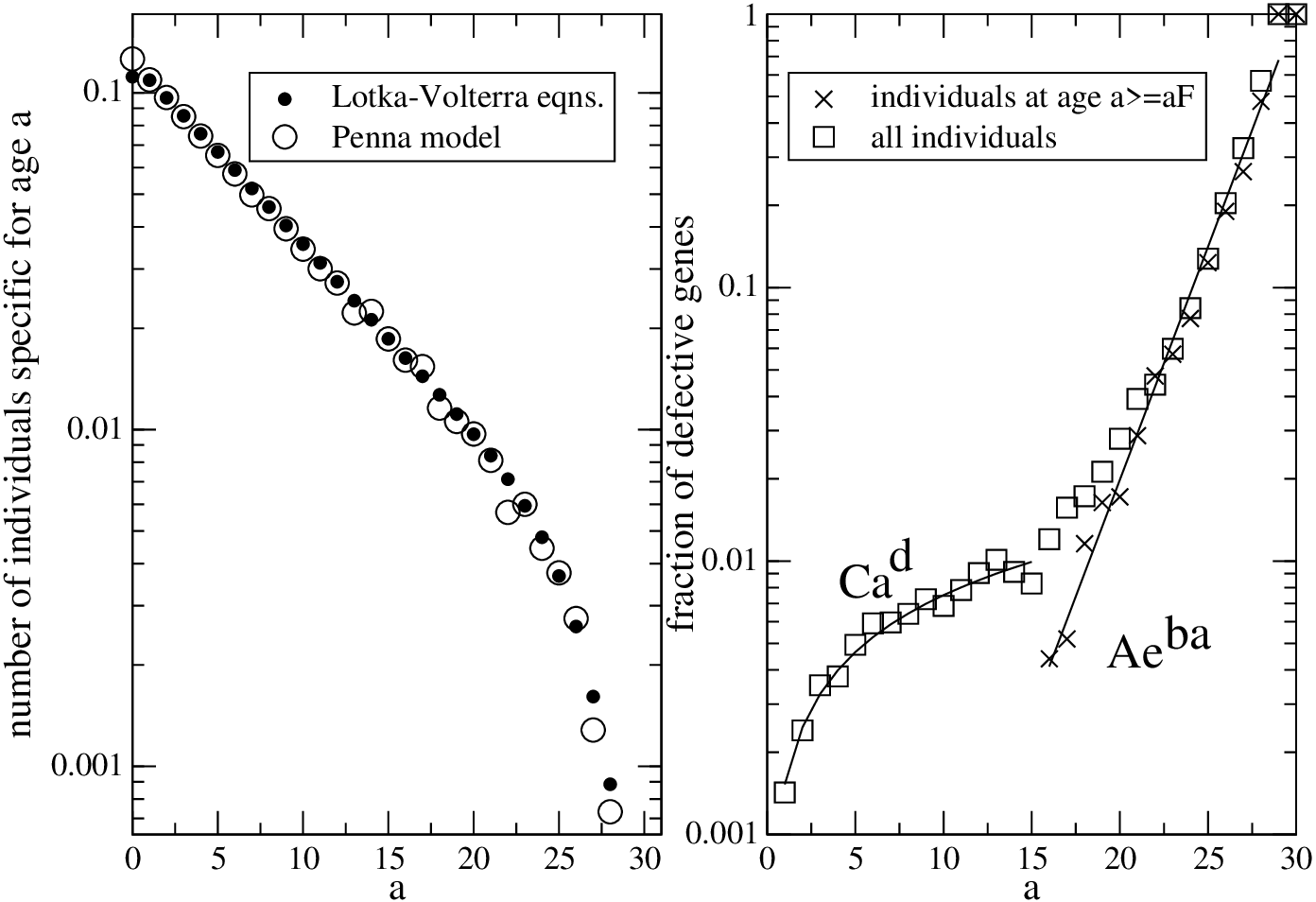}
\end{center}
\caption{The dependence of the population number on the individual's
  age in the case of the equilibrium Penna model of the bit-string
  evolution ($T=1$,$a_F=15$, $M=1$) and the age-structured
  Lotka-Volterra equations (the
  left panel of the figure). The right 
  panel  shows the distribution of defective genes in the
  equilibrium Penna model for the whole population and for individuals of age $\ge
  a_F$. The solid lines represent an approximation of the data $\varepsilon_a$ (fraction of defective genes in the population) which is power-like for the ages $a\le a_F$ and exponential-like for  $a > aF$.
}
\label{fig2}
\end{figure}

Thousands of papers have been published on the Lotka-Volterra
  equations \cite{lotka1,lotka2} describing population growth, 
competition or speciation. In real populations the reproduction rate of individuals 
depends on their age and therefore it is necessary to include age structure into these equations. One example of how this can be achieved can be
found in \cite{lotka-age}. It is also possible to introduce
a time delay between cause and effect (see,
e.g. \cite{lotka-delay}). However, the majority of the Lotka-Volterra
equations do not usually include 
genetic information and the question arises how  to include it
directly into the Lotka-Volterra equations.

The age-specific equations for population growth seem to be a good candidate to represent genetic information because the age structure introduces some analogy to the Penna model \cite{Penna1,Penna2} of genetic evolution. 
It is a model of genetic evolution although all details concerning the
genes are skipped except of the state of their functionality - is the gene under
consideration correct or it is mutated. This simple model 
has turned out to be very successful in interpreting  the demographers data of real populations even as complex as human populations \cite{Penna2,Stauffer1,demographer,Wroclaw}.

In the original asexual version of 
the Penna model \cite{Penna1}, the population under consideration consists of 
individuals represented by genomes defined as 
a string of $n$ bits. The bits represent states of genes  where  $0$
 denotes its functional allele and  $1$ its
 bad allele. It is assumed that if an individual possesses $T$ bad alleles switched on, it dies. 
In the model all genes are switched on chronologically - each bit 
corresponds to one
"year" - and maximum life span of an individual is $a_D$ "years". 
After reaching the fertility age $a_{F}$ an individual gives birth to $B$
  offsprings whose genomes are mutated versions of the parental genome. The mutation rate 
$M$ is constant. After the mutation the affected gene is represented by a bit with a value opposite to the value before the mutation. The results of the Monte Carlo simulations of three different haploid populations for the Penna model in the case when back mutations from 1 to zero are not allowed have been shown in  Fig.~\ref{fig1}. In the figure, there have been plotted fraction of defective chronological genes in each population and age distribution of individuals.

In the diploid version of the Penna model the individual's genome is
represented by two bitstrings and then each locus possesses two
alleles. The diploid individuals can reproduce sexually
\cite{Penna2}. Both the haploid model of the equilibrium population
and its diploid version are uniquely described by the
fraction of defective genes in the population specific for each individual's
age. A short review of Monte Carlo simulation results for the Penna model with some
  additional details, like the presence of the housekeeping genes or
  the recombination frequency, can be found in \cite{Wroclaw}.

Although the age distribution curves obtained in the Penna model
coincide very well with demographers data for real populations, the
model is not so ``interdisciplinary'' as the Lotka-Volterra population
model. The reason could be that it is very difficult to obtain
analytical results for the general Penna model and therefore one has 
to use the Monte Carlo method. However Monte Carlo simulations of 
the Penna model need large populations and the simulation time is very
  long. Hence, the typical multi-species problem  exceeds the
computing capability of a single
PC-computer. Another problem is how to avoid the correlations 
arising from parallel computing if one tries to distribute
  the simulations to many processors. 
On the other side hand, it is relatively easy to solve
numerically even a large set of the differential equations describing the Lotka-Volterra populations. Below we show how to include the fraction of defective genes
  into the equation for population growth so that the age distribution
  curves for these two models coincide.

In the following sections we discuss the behavior of the
Lotka-Volterra ecosystem in which the age-specific species has been
determined by the form of the fraction of defective genes. 

\section{Population growth of a single species}

In this section we restrict ourselves to the haploid version
of the Penna model  but the results could be generalized to diploid
populations. All we need from the Penna model
is the distribution of defective genes in an equilibrium population and 
the corresponding age distribution. Such data could be taken from a real 
population as well. In the case of the Penna model we have performed a series of
simulations of genetic evolution of the bitstring populations for
different parameters like the fertility age $a_F$, genome length and
the number of offsprings born each year. We
considered the simplified case when  
the parameter $T=1$. The Verhulst factor was used
to control the population size. 

In the right panel  of the
Fig.~\ref{fig2} in the semi-log scale  
there have been plotted separately the fractions $\varepsilon_a$ of
the age-specific defective genes in the whole equilibrium population
and in the part of it which is consisting of the individuals with the
age $a \ge a_F$. Fraction $\varepsilon_a$ of
the age-specific defective genes in the whole equilibrium population
and in the part consisting of individuals with $a \ge a_F$ are plotted
in Fig.~\ref{fig2} on semilog axes.
 In the latter case all individuals should posses good
genes specific for $a \le a_F$ ($\varepsilon_a=0$) because $T=1$. Otherwise they should
have died. 

Our deterministic model for population growth is constructed in such a
way that the values of the fractions $\varepsilon_a$ for each age
$a=0, 1, \ldots, a_D$  are inserted into the following set of the
age-specific differential equations describing the change in number of
individuals of age $a$:

\begin{equation}
\frac{dN_a(t)}{dt}=\alpha_{a-1}  V(t) (1-\varepsilon_{a-1}) N_{a-1}(t) -  \gamma_a N_a(t),
\label{lotkasingle}
\end{equation}

\noindent
where $N_a$ is the number of individuals specific for the age $a=1, \ldots, a_D$

\begin{equation}
 N_0=\sum_{a \ge aF} N_a(t).
\label{N0}
\end{equation}

\noindent
and $V$ represents the Verhulst saturation factor
\begin{equation}
 V(t)=1-\frac{1}{\Omega} \sum_{a=1}^{a_D} N_a(t).
\end{equation}

\noindent 
with $\Omega$ being the saturation level. The first term is the
graduation term from the age $a-1$ to the age $a$ whereas the other
part of the equation describes the graduation to the age $a+1$ and the
individual's mortality at age $a$. These terms are controlled by the
age-specific rate coefficients $\alpha_a$ and
$\gamma_a$, respectively. If we restrict the above to genetic evolution only, 
as in the case of the Penna model, we would have to  choose the values
$\gamma_a=1$ for all ages $a$, the reproduction rate $\alpha_0$ should
be chosen as in the computer simulations of the Penna model (e.g. we set it to 1.1) 
but $\alpha_a=1$ for all  ages $a>0$.  
The Verhulst factor (the same as in the Penna model) ensures that the
solution of the equations Eq.(\ref{lotkasingle})  saturates at long
times and then the age profile in the population coincides
with the one from the equilibrium Penna population. An example of
this can be observed in the left part of Fig.~\ref{fig2},
where an analytical 
approximation of the simulation data $\{\varepsilon_a\}_1^{a_D}$ has
been applied.

Equations Eq.(\ref{lotkasingle}) seem to possess many parameters. 
However, the simulations of the simple haploid 
 version of the Penna model for different values of $a_F$ suggest that
 in the case of $T=1$ the fraction of the age-specific defective genes
 in the population which can be activated, i.e. after which the individuals die, 
 consists of two parts (Fig.~\ref{fig2}), one part relates  to individuals of age $a \ge a_F$ and another one relates to individuals of age $a < a_F$. In this simple case  we
 can approximate them as follows: 

\begin{equation}
\varepsilon_a \sim C a^{d}
\label{branch1}
\end{equation}

\noindent
for $a \le a_F$ and 

\begin{equation}
 \varepsilon_a \sim A e^{ b a} 
\label{branch2}
\end{equation}

\noindent
for $a>a_F$, with  constants $A$, $b$, $C$, $d$. 
From the normalization condition, $\varepsilon_{a_D}=1$, we obtain 

\begin{equation}
A=e^{-b a_D}.
\end{equation}

\noindent
Then, the number of parameters describing the fractions of
defective genes drops down. 
It can be even smaller if we  
continue the left branch 
of the function $\varepsilon_a$ to $a=a_F+1$ and merge it with the
right branch at $a=a_F+1$  in such a way that individuals of age $a \ge a_F$ did not contributed 
into the left branch also at $a=a_F+1$, i.e.

\begin{equation}
C (1+a_F)^d - A e^{1+aF} = A e^{1+aF}.
\end{equation}

\noindent
Then we could estimate the
value of the parameter $C$: 

\begin{equation}
C=2 A e^{b (1+a_F)}/(1+a_F)^d.
\end{equation}

\noindent 
In this way we have got an approximate distribution of defective genes
controlled by  four parameters, $b$, $d$, $a_F$ and $a_D$.  In the
equations Eq.(\ref{lotkasingle}) there is also the
parameter $\alpha_0$, but it was kept constant and its value was the
same as in the computer simulations of the Penna model.

The small number of parameters in the deterministic model makes it possible to consider
multi-species evolution where the species 
differ in the values of $a_F$, $a_D$, $b$, $d$ and
$\alpha_0$. In this case the numerical solutions do not require a high computational
effort as in the Penna model. We should add that if the distribution function of
the defective genes is an unknown function of $a$ and only the empirical
values $\varepsilon_a$ are available, then these values could be used in
the same way as in the discussed example, i.e., we do not need the
analytical form of the distribution function for defective genes. We
have considered only the case of the genetic evolution in
equations Eq.(\ref{lotkasingle}) but there is a rich
possibility to use other values of the coefficients  
$\alpha_a$ and $\gamma_a$ if one wants to describe real populations.

\section{Predator-prey system}

We consider an ecosystem in which a limited amount of the
self-regenerating energy resources, $S(t)$, are consumed by the
age-structured species living in this ecosystem. In this case,  
the increase in number of the individuals in the species depends on
the available energy resources necessary for life processes. We will
restrict ourselves to one species only.   The species will carry
genetic information represented by the fraction of defective genes in
the population, the fertility age and the longevity $a_D$. We will
show that the change of these genetic parameters can 
change the trend toward equilibrium of the whole ecosystem.

In the case of one species the ecosystem can be described by the
following predator-prey set of equations

\begin{eqnarray}
\frac{dS(t)}{dt}\,\,\, & = & \alpha_S S(t)\,(1 - \frac{S(t)}{\Omega_S}) - \gamma_S N(t) S(t) \nonumber \\
\frac{dN_a(t)}{dt} & = & \alpha_{a-1}  S(t) (1-\varepsilon_{a-1}) N_{a-1}(t) - N_a(t), \nonumber \\
\label{Ecosystem}
\end{eqnarray}

\noindent
where $a=1, \ldots, a_D$, $N_0$ is the same as in Eq.(\ref{N0}),
$N(t)$ is the total number of individuals 

\begin{equation}
 N(t)=\sum_{a=1}^{a_D} N_a(t),
\end{equation}

\noindent
$S(t)$ represents energy units consumed by the species, $\Omega_S$
represents the saturation level for the energy resources,  
the coefficient $\alpha_S$ is the regeneration rate of the ecosystem
resources, $\gamma_S$ is the damping coefficient due to the species
using the energy resources. In order to follow the Penna model of
genetic evolution the species evolves according to the   
equations Eq.(\ref{lotkasingle}) ($\gamma_a=1$ for all values of $a$)
but the species' growth is controlled by the amount $S(t)$ of the
available energy resources instead of the Verhulst factor,
$V(t)$. However, a Verhulst term is assumed for the
self-regeneration of the ecosystem with respect to the energy resources. 
The energy resources represent prey and the species plays the role of
predator. 
One could easily generalize the above set of equations to include many
species. 

We could expect that equations
Eq.(\ref{Ecosystem}) have solutions 
which change in time in a similar way as the
microscopic Penna model 
evolution in a limited ecosystem. We
  have recently studied an analogous predator - prey problem in
  \cite{dudek1}, where  a variable surrounding and its effect on the
  Penna model was considered. One of the 
findings of the computer simulations was that small changes in the
inherited genetic information could lead to spontaneous bursts of
evolutionary activity. The predator-prey dynamics with genetics could
be as complex as discussed by Ray et al. \cite{Ray} who showed that
the system passes from the oscillatory solution of the Lotka-Volterra
equations into a steady-state regime, which exhibits some features of
self-organized criticality (SOC). 

\begin{figure}
\begin{center}
\includegraphics[scale=0.4]{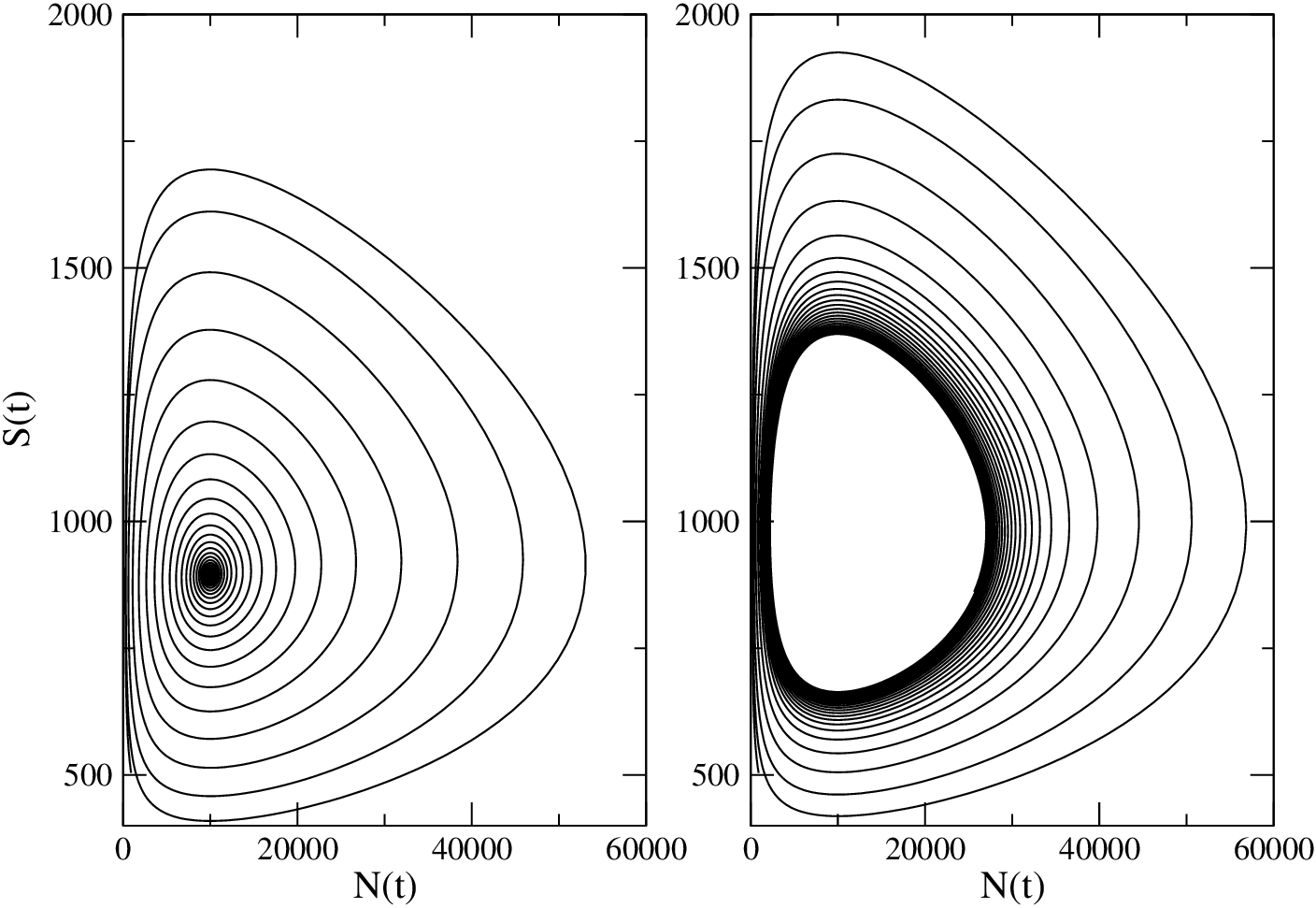}
\end{center}
\caption{Both panels show
  the variation of the energy resources $S(t)$ and the total number of
  individuals $N(t)$ of the age structured species starting from the
  same initial conditions.  The difference is that on the left $a_F=15$
  and on the right $a_F=20$. The remaining parameters are the same,
  $a_D=29$ and $\alpha_0=1.1$.  
}
\label{fig3}
\end{figure}

In our model, we can always expect saturation
of the energy resources if there is no species ($N(t)$=0) because the
Verhulst term is present in Eq.(\ref{Ecosystem}). However, if $N(t)>0$
then depending on the shape of the distribution function of defective
genes the solution of Eq.(\ref{Ecosystem}) saturates as in the example
in the left part of  Fig.~\ref{fig3} or
it oscillates as in the 
right part of the figure. This means that an ecosystem with a few
species can exhibit very complex evolution in which the solutions
representing some species will spiral towards an equilibrium fixed point 
and some of them will try to converge to the limit cycle. 
The smaller the value of $a_F$ is, the larger the oscillations are, and the
  species may become extinct. 

It is important that not only genetic parameters can change the type
of the asymptotic solutions of Eq.(\ref{Ecosystem}). Consider a species 
similar to the one represented in the left hand part of, Fig.~\ref{fig3} for
  which the Lotka-Volterra solutions spiral to a fixed point. Let's
  increase the value of the fertility age from $a_F$ to $a_F^{\prime}
  > a_F$ after some period of time but let the genetic information
  represented by the values $\{ \varepsilon_a\}_1^{a_D}$ remain the
  same as before. For example, according to some social
regulations since a specific time moment the individuals start to
reproduce  at older ages $a \ge
  a_F{\prime=20}$.  The consequences of this shift in the reproduction age is that the
solutions of Eq.(\ref{Ecosystem}) change qualitatively from the trend
approaching the fixed point as in the left part of Fig.~\ref{fig3}
to oscillations as in the case of the species from the right hand
part of  Fig.~\ref{fig3}. It could
happen that such a transition 
between these two types of  solutions of the
Lotka-Volterra equations could cause the species to become extinct.

\begin{figure}
\begin{center}
\includegraphics[scale=0.4]{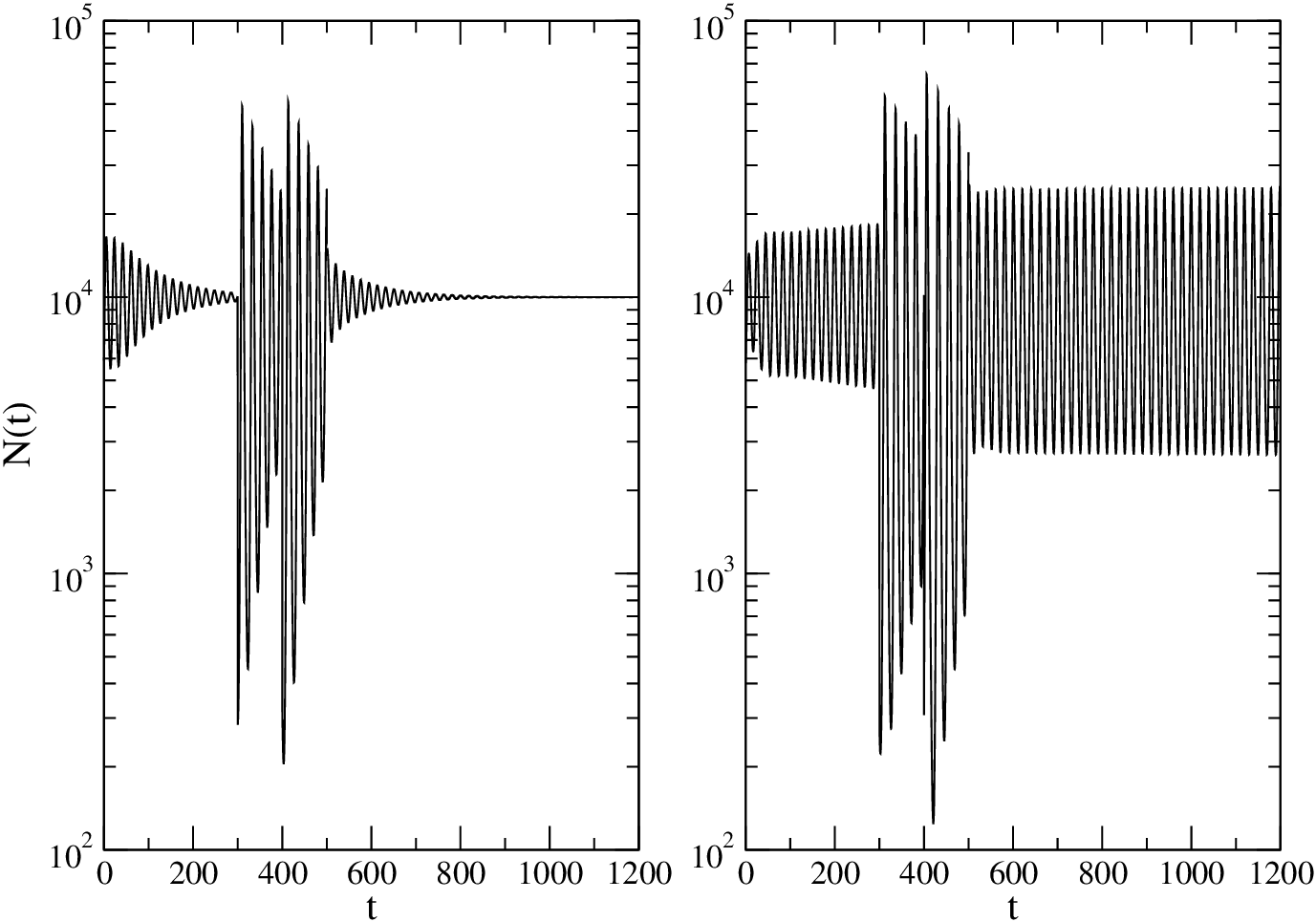}
\end{center}
\caption{The effect of three catastrophes, where $30\%$ of the population under
  consideration has been eliminated at two time moments $t=300$ and
  $400$ and $10\%$  of the population at $t=500$. On the left there is
 a species with $a_F=15$ and $a_D=29$ and on the right the
  species has $a_F=20$ and $a_D=29$. In both cases the catastrophe  at
  $t=500$ decreased the population number fluctuations.}
\label{fig4}
\end{figure}

\begin{figure}
\begin{center}
\includegraphics[scale=0.3]{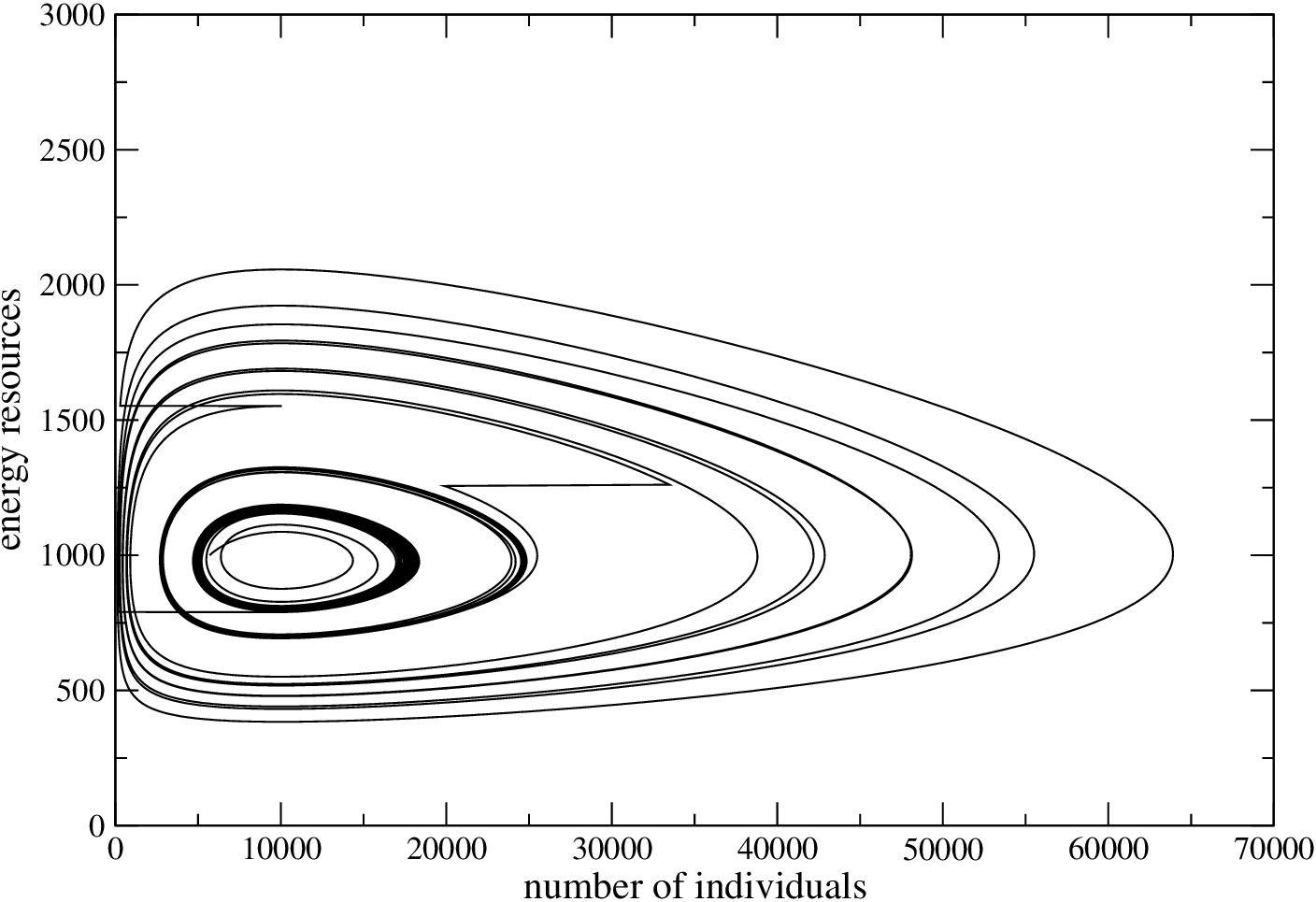}
\end{center}
\caption{The effect of catastrophe from Fig.~\ref{fig4} on the variation of  $S(t)$ and $N(t)$ in the population with $a_F=20$ and $a_D=29$.}
\label{fig5}
\end{figure}

In \cite{StaufferPekalski} it 
has been shown that the Lotka-Volterra systems have a self-regulatory
character and there exist threshold values for the fraction of
destroyed population above which the system returns to its previous
state. We observe a
  similar behavior in our model. In particular, in
Fig.~\ref{fig4} an effect of three
  such catastrophes on two different species represented by two
  different types of solutions of the Lotka-Volterra equations is
  shown. It is interesting that for some solutions the
  introduction of a perturbation decreases the amplitude of their
  oscillations. This could be observed in  Fig.~\ref{fig4}. 
The trace of this catastrophe in the $(S(t),N(t))$ representation has been shown in Fig.~\ref{fig5}.

\begin{figure}
\begin{center}
\includegraphics[scale=0.4]{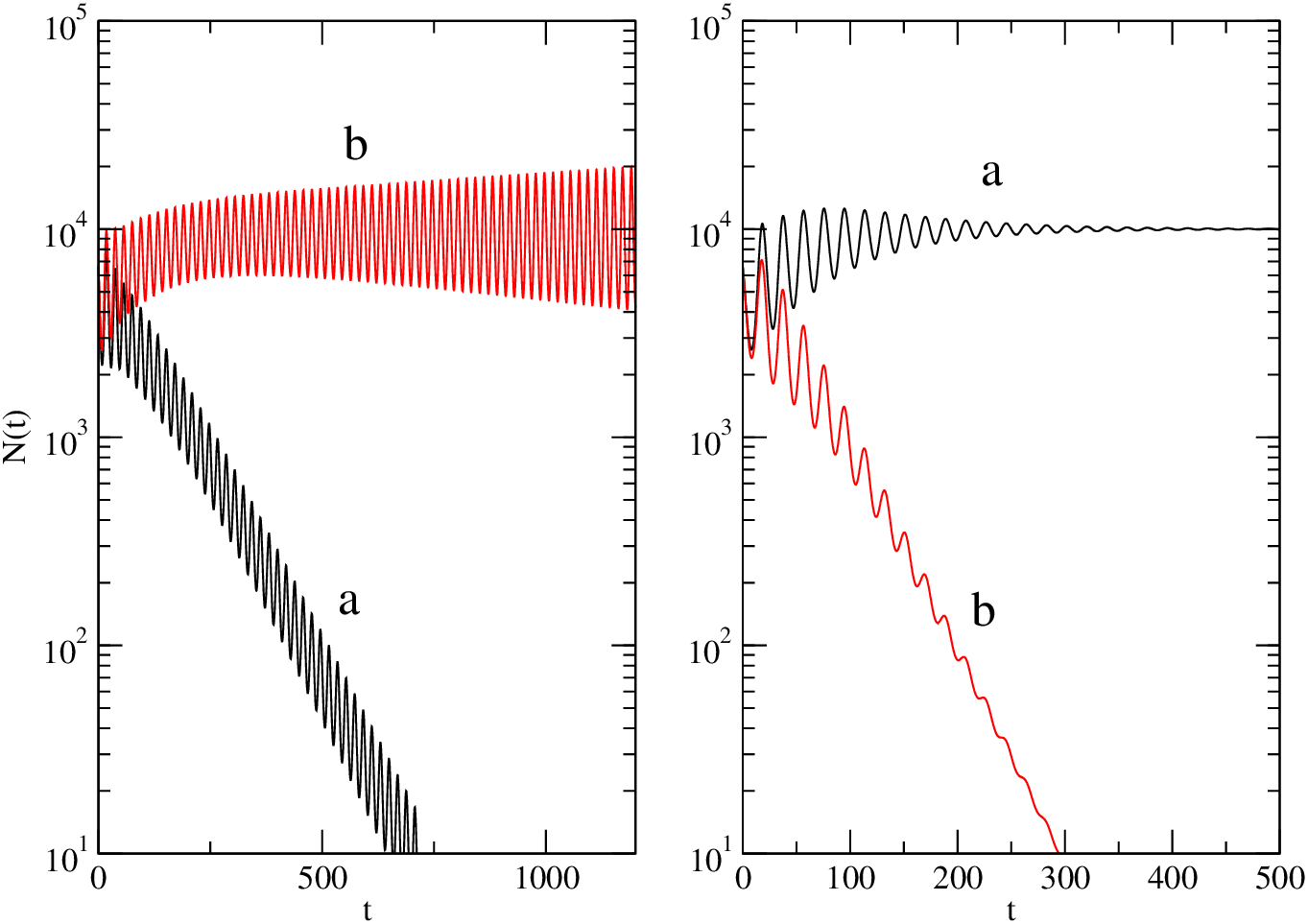}
\end{center}
\caption{Dependence of population number $N(t)$ on time $t$ in the
  case of two species (a) and (b) competing for the same energy
  resources. In both figures, the species (a) has the same fertility
  age $a_F=15$ but $a_D=23$ on the left and $a_D=24$ on the right.
  The species (b) ($a_F=20$ and $a_D=29$) is the same in both
  cases. The same initial condition have been applied.} 
\label{fig6}
\end{figure}

Thus, we could expect that in the ecosystem consisting of many species
a rapid change like the extinction of a species or a sudden increase in their 
number can cause discontinuous changes of the 
fluctuations of the population numbers of the existing species, similarly as in 
Fig.~\ref{fig5} or Fig.~\ref{fig6} (discussed below). 
The age-specific structure of the competing populations seems to  be
an important factor in their survival.

It is easy to generalize the Lotka-Volterra equations
Eq.(\ref{Ecosystem}) to the case  
of many species competing for the resources. If index $k$ runs through $M$
  different species,
the age-structured growth equations should be changed to the following:

\begin{equation}
\frac{dN_a^k(t)}{dt} =  \alpha_{a-1}^k  S(t) (1-\varepsilon_{a-1}^k) N_{a-1}^k(t) - N_a^k(t), \end{equation}

\noindent
where $k=1, 2, \ldots, M$. Even in the case of two species $M=2$ the
ecosystem under consideration can exhibit complex behaviour. For example, Fig.~\ref{fig6} shows the time dependence
of the population number $N(t)$ for two competing species (a)
and (b) in two cases which differ for species (a) by one parameter only, $a_D=23$ (left) and
$a_D=24$ (right).
The shift in the value of $a_D$ decides which species may become extinct.

\section{Conclusions}
It has been shown how to construct 
age-structured Lotka-Volterra 
equations in order to  describe some features of genetic evolution. 
This could be helpful in modeling real populations
where many parameters
  are typically used. In the considered model all species were
competing for the same energy resources. 

\section*{Acknowledgement}
Autor thanks D. Stauffer for discussion and helpful suggestions.

\end{document}